\documentclass[a4paper,fleqn]{cas-sc}

\usepackage[authoryear,longnamesfirst]{natbib}

\usepackage{xcolor}
\usepackage{graphicx,url}
\usepackage{multirow}
\usepackage{float}
\usepackage{makecell}
\usepackage{amsmath}
\usepackage{subfigure}
\usepackage{makecell}
\usepackage{longtable}
\usepackage{adjustbox}
\usepackage{booktabs}
\usepackage{fixltx2e}
\usepackage{lineno}
\usepackage{comment}
\usepackage{rotating}
\usepackage{tablefootnote}
\usepackage{fontawesome}
\usepackage{tabularx}
\usepackage{makecell}

\usepackage{dashrule}
\usepackage{tikz}
\usetikzlibrary{decorations.markings}

\usepackage{fontawesome}

\usepackage[linesnumbered,ruled,vlined]{algorithm2e}

\usepackage{pifont}
\usepackage{float}

\def\tsc#1{\csdef{#1}{\textsc{\lowercase{#1}}\xspace}}
\tsc{WGM}
\tsc{QE}
\tsc{EP}
\tsc{PMS}
\tsc{BEC}
\tsc{DE}

\begin{document}
\let\WriteBookmarks\relax
\def\floatpagepagefraction{1}
\def\textpagefraction{.001}

    \shorttitle{Qualitative Assessment of Low Power Wide Area Network Protocols and their Security Aspects}


\title [mode = title]{Qualitative Assessment of Low Power Wide Area Network Protocols and their Security Aspects}                      



%

\author[1,3]{WR Bezerra}[type=editor, orcid=0000-0002-6098-7172]
\cormark[2]
\fnmark[1]
\ead{wesleybez@gmail.com}

\author[1,3]{LM Bezerra}[type=editor, orcid=0000-0003-3331-7277]
\cormark[2]
\fnmark[1]
\ead{lais.machado1@gmail.com}

\author[1]{Carlos B. Westphall}[type=editor,orcid=0002-5391-7942]
\cormark[1]
\ead{carlosbwestphall@gmail.com}

\address[1]{UFSC - Federal University of Santa Catarina,Campus Universitário - Trindade,Florianópolis/SC - Brazil, 88040-380}

\address[1]{Independet Researcher - Aurora/SC - Brazil, 89186-000}

\cortext[cor1]{Corresponding author}
\cortext[cor2]{Principal corresponding author}



\begin{abstract}
   There are currently many communication options in the Internet of Things, even in particular areas such as constrained and battery-powered devices, such as Low Power Wide Area Networks. Understanding the differences and characteristics of each option is a challenge, even for professionals and researchers in the field. To meet this need, this work analyses the qualitative characteristics of Low Power Wide Area Network protocols and the challenges and opportunities of using constrained devices for sparse networks based on long-life batteries. For this study, a bibliographic survey of the literature was carried out as an analysis of three protocols (LoRaWAN, NB-IoT, and Sigfox), and a detailing of the first one. As a result, there is a discussion about the chosen network protocol and its use in IoT solutions with sparse sensors.
\end{abstract}


\maketitle

\section{Introduction}
\label{sec_introduction}

This article provides a list of low-power long-distance (LPWAN) protocols most commonly used in remote monitoring solutions with sparse sensors that have low processing and memory capacity (Constrained Devices - CD) \cite{elhanashi2024advancements, nilima2024optimizing} which use ultra-long-life batteries (ULLB). This type of equipment poses challenges for providing services with a certain level of security. For example, due to cost limitations, some more modest processors do not provide the precision required to implement encryption solutions \cite{singh2024advanced, kumar2024review}.

Nowadays, there is a wide range of options for which type of long-distance network to use in this system. These network options provide connectivity to remote environments. Among them, some data communication modes stand out, such as low-orbit satellite \cite{ortigueira2024satellite, ahmed2024integrated, shayea2024integration}, non-terrestrial networks \cite{simo2021air,mejia2025development,park2018forestry} (i.e., drones), vehicular networks \cite{sharma2024lorawan,twahirwa2021design,klaina2020aggregator}, cellular networks \cite{abbas2024towards, uzoka2024role}, all of which integrate with LPWAN networks, which in our perspective is still the most appropriate type of network for this situation. However, it is important to know more about the characteristics of LPWAN networks and evaluate their performance in situations similar to our case study.

The proposed scenario does not have a high density of sensors, but instead sparsely distributed sensors, which have their energy supply based on batteries and do not need to transmit/receive large amounts of data \cite{olsson20146lowpan,routray2019narrowband}. This usage scenario has characteristics such as the need for restriction in using computational resources since these networks do not have many resources and do not transmit large amounts of data. It is observed that there is a low need for data transmission. In this situation, LPWAN networks are more suitable and have been in use for some time \cite{mekki2019comparative,routray2019narrowband,khan2018iot}. Even with the evolution of other data transmission forms, this type of network is still essential and current.

This work talks about LPWAN networks of cellular and non-cellular types. Qualitative issues and aspects of this type of network are raised, such as transmission quality and maximum distance. To evaluate the most used protocols, after this evaluation, the protocol will be selected, and this protocol will be further detailed. This details the architecture of this protocol and how it can be best used to solve a use case where forest monitoring is used specifically fire and forest monitoring. Specifically, we can list the contributions as follows:

\begin{itemize}
\item a survey of the challenges and opportunities of using LPWAN and CD for monitoring;
\item a qualitative analysis of the characteristics of the LPWAN protocols: LoRaWAN, NB-IoT, and Sigfox;
\item a more detailed evaluation of the LoRaWAN protocol; \end{itemize}

The remainder of the document is organized as follows: Section \ref{sec_methodology} is followed by Section \ref{sec_desafios_lpwan_iot} with a characterization of the opportunities and challenges of these technologies for environmental monitoring. Section \ref{sec_analysis} analyzes the selected protocols and details the LPWAN protocol. Section \ref{sec_discussion} analyzes the results found. Finally, Section \ref{sec_conclusion} presents the conclusions and future work.

\section{Methodology} 
\label{sec_methodology}

\begin{figure}[h]
\centering
\includegraphics[width=0.75\linewidth]{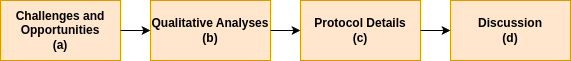}
\caption{Work methodology}
\label{fig:methodology}
\end{figure}

This study was carried out in four stages: \textbf{(a)} an analysis of challenges and opportunities, \textbf{(b)} a qualitative analysis of LPWAN protocols, \textbf{(c)} a detailed analysis of the selected protocol, and \textbf{(d)} a discussion of the selected protocol. The methodology is expressed in Figure \ref{fig:methodology}.

The \textbf{analysis of challenges and opportunities (a)} was carried out through a survey of bibliographies related to the monitoring of sparse environments such as forests, plantations, and similar situations. The LPWAN protocol analysis (b) is based on four characteristics (characteristics, strengths, weaknesses, and open issues) combined with a summary in a comparative table and a table with threats cited for each protocol.

Continuing with the protocol detailing (c), LoRaWAN was selected for detailing. This part presents the protocol architecture, its device characteristics, and security issues. Finally, the discussion (d) presents and justifies the motivation for choosing this protocol for our usage scenario.

In summary, this methodology presents the scenario through challenges and opportunities, followed by more general facts about the protocols through qualitative questions evaluated through existing studies. It consequently evolves to a more detailed evaluation of the LoRaWAN protocol. It closes by arguing how the characteristics of the selected protocol stand out concerning the others given the application case of this data transmission technology.

\section{Challenges and Oportunities in CD and LPWAN}
\label{sec_desafios_lpwan_iot}

Internet of things is an emerging technology, been researched remarkably for over a decade \cite{deep2022survey}. Such technology integrates different types of equipment with connectivity to the internet. Generally, it offers a platform for data crossing, creation of usage profiles, and intelligence extraction from the obtained data. Its adoption and growth have been striking; it has been cited as the internet of the future, according to authors such as Yi \textit{et al}. \cite{yi2015fog}, However, that should be seen as both an opportunity and a threat.

The opportunity comes from new services, equipment, and amenities available to users at a more affordable and user-friendly price. Nevertheless, the threat is related to the amount of equipment, many of which are \textbf{poorly configured}, and the increasing computational power of malicious users. Furthermore, a \textbf{massive denial of service attacks}, \textbf{insecure network services}, \textbf{interfaces of cloud/mobile/web}; are some of the examples \cite{miessler2015securing} of problems arising from equipment that implements poor security or is poorly configured.

In addition, the IoT is present in several areas, such as industry, home automation, building automation, home security and surveillance, Smart Cities, and Smart Agriculture. Hence, such technology has proven to be a key factor for \textbf{better performance in producing goods and food}, and improving the \textbf{quality of life} of people worldwide. As a result, it is a promoter of \textbf{better agricultural performance}, IoT has been used in several agricultural areas, enabling financial benefits and compliance with legislation.

Specifically, some of the agriculture areas which have benefited from using IoT include oyster farming, shrimp farming, fish farming, and precision agriculture. Comparatively, solutions contributing to the effective use of agricultural inputs have \textbf{lowered costs}, \textbf{saved water}, and \textbf{decreased pesticide use}. So, with more information about the soil, it is possible to make corrections and decrease the effects of pesticides, soil wear, and water contamination. So far, another important application is in the \textbf{water resources use monitoring}. In such manner, it is possible to preserve it, \textbf{reducing pumping costs} and \textbf{preventing soil leaching}.

Due to its large entry into several sectors, IoT is a promising reality and a challenge regarding security issues. This dilemma is linked to the \textbf{emergence of this technology} and the fact that there are \textbf{still no well-established security mechanisms} for such an application, according to Tiburski \textit {et al.} \cite{tiburski2019lightweight}. IoT has been applied in several sectors, from areas that do not have an immediate connection with technology to areas that use cutting-edge technologies in their daily functions. Even though such area often has professionals more qualified to implement such mechanisms, family farming does have.

Each IoT system \textbf{solution involves several parameters} thereby lacking of user technical experience can influence the poor security mechanisms adoption. Parameters could include the data link, the reliability required for data transmission, time constraints, and delays. Additionally, another important factor is the ability to \textbf{supply energy} to the devices involved. While some devices will be directly connected to the power supply grid, others will use batteries or harvest energy (\textit {i.e.}, Solar); for data processing and transmission. This becomes a central factor in a low-cost device. Since such devices were designed with energy restrictions, all their processes must use as less energy as possible $-$ this also concerns safety.

\begin{table}[h]
    \centering
    \caption{Opportunities and Challenges for Constrained Devices IoT}
    \begin{tabularx}{\textwidth}{@{}XX@{}}
        \toprule
        \textbf{Challenges \faArrowCircleODown}  &  \textbf{Opportunities \faArrowCircleOUp} \\
        \toprule
        Poorly configured devices & Better performance producing goods and food\\
        Insecure network services & Quality of life\\
        Insecure interfaces cloud/mobile/web & Lowered costs\\
        Yet an emergent technology & Water saving\\
        Not well established security mechanisms & Decreasing pesticide use\\
        Solution involves several parameters & Monitoring water resource use\\
        Energy supply/source & Reducing pumping costs\\
        & Preventing soil leaching\\
        \bottomrule
    \end{tabularx}
    \label{tab:challenges_opportunities}
\end{table}

Also, that is a focus on constrained devices, which run on batteries, as well as having a radio link (wireless) with a low capacity for transmitting data over long distances (LPWAN). In the forest monitoring deployment, this type of devices will be sparsely distributed and potentially isolated geographically. Such low cost devices, based on cheaper hardware and more modest in terms of resources, is what allows the dissemination of IoT in areas with less investment in technology in developing countries such as agriculture \cite{mathe2015local,akinwole2007biological,gabriel2007locally}, irrigation \cite{bueno2020improving, bueno2020improving}, water resources monitoring \cite{ugwuanyi2021survey,wu2019lora,liu2018solar,prasad2015smart}, among others. Such IoT areas have less investment than areas such as Industry 4.0; however, they greatly benefit the population with lower social conditions.

\section{Protocol Analysis}
\label{sec_analysis}
LPWAN has a lower transmission rate, in consequence there is less power consumption, being an optimal solution for constrained IoT devices. Such devices depend on batteries and does not need to frequently transmit large amounts of data. In other words, even with a low baud rate, such protocols can provide the necessary connectivity and reach long communication distances and also are suitable for agriculture and forest monitoring solutions.

In brief, several protocol options on the market can be classified in different ways, such as the type of band and infrastructure. As for the type of band, it can be licensed or unlicensed. As for the type of infrastructure, it can be cellular or non-cellular. These characteristics classify all LPWAN protocols, even the ones that are less expressive such as Ingenu \cite{almuhaya2022survey}, Weightless \cite{laya2016goodbye}, LTE-M \cite{abou2021nb}, DASH7 \cite{ayoub2018internet}, etc., as the most commonly used LPWAN protocols \cite{islam2024future} such LoRaWAN, NB-IoT, and Sigfox \textemdash  these three are detailed next in the text.

\begin{table}[h]
    \centering
    \caption{Comparison between LPWAN technologies.}
        \label{tab:lpwan_comparative_qualitative}
    \begin{tabularx}{.8\textwidth}{@{}XXXX@{}}
    \toprule
        \textbf{\#} & \textbf{LoRaWAN} & \textbf{NB-IoT} & \textbf{Sigfox}  \\
    \midrule
        \textbf{Origin} &France &3GPP &France \\\hline
        \textbf{Creation Year} &2015&2015/2016&2010\\\hline
        
        \textbf{Security} & AES-128& AES-128& AES-128\\\hline
        \textbf{Band} &unlicensed &licenced & unlicensed\\\hline
        \textbf{Modulation} &CSS/BPSK&\makecell[l]{SC-FDMA\\OFDMA}& DBPKS and GSFK \cite{kadusic2022smart}\\\hline
        \textbf{Data rate} &0.3 - 5.5kbps&\makecell[l]{(UL)200kbp\\(DL)20kbps}&100bps\\\hline
        \textbf{Coverage (rural)} & $<$ 20km&$<$10km&$>$40km\\\hline
        \textbf{Coverage (urban)} & $<$ 5km&$<$1km&$<$10km\\\hline
        \textbf{Bandwidth} &125/250/500kHz \cite{coman2019security}&200Hz&100Hz\\  \hline      
        \textbf{Max Payload} &243 bytes &1600 bytes&12 bytes\\\hline
        \textbf{Topology} & Star& Star& Star\\\hline
        \textbf{Max End Devices} &50k&100k&50k\\\hline
        \textbf{Duty cycle} &01\% \cite{benkahla2018enhanced}&no limit&01\%\\
    \bottomrule
    \end{tabularx}

\end{table}

\textbf{LoRaWAN} is an emerging technology, introduced in 2015 \cite{khalifeh2021lorawan}, that allows systems to transmit a small data rate over large distances and at a low-cost \cite{lin2017using}. Once LoRa has no restrictions on its deployment, an interested company or person could implement a LoRaWAN system without restrictions. Further, it was initially developed by French startup Cycleo and sold later to American Semtech \cite{mekki2018overview}. Besides LoRa being a proprietary technology, LoRaWAN has great acceptance by developers due to its openness and open source-related solutions as the gateway.

As for \textbf{network}, such protocol is based on Semtech's LoRa technology and uses Chirp Spread Spectrum (CSS) modulation \cite{thaenkaewevaluating}. Also, this technology has transmission capacities (data rate) from 0.3kbps to 50kbps, depending on the required distance and interference to which the signal is subject. However, it has enough range to cover a city using only three base stations \cite{mekki2019comparative} what is a comparatively long coverage. Specifically, LoRa defines ten communication channels, eight of which are multi-data (250 bps - 5.5 kbps), a single data rate channel (11 kbps), and a Frequency Shift Keying (FSK) channel (50 kbps) \cite{gresak2019detection}. Consequently, this protocol presents some advantages and disadvantages inherent to these network characteristics.

Its \textbf{main positive characteristics} are security, battery life, multiple communication channels, and different classes of \cite{naoui2016enhancing} devices. Moreover, it was designed for Ultra-Long Life Battery (ULLB) devices and can meet the requirements of different applications. Also, it is a protocol that provides means of adding new nodes to the network in a secure way \cite{stefano2017security} \textemdash, in the v1.1 version of the protocol. Furthermore, it has many advantages such as good coverage, is suitable for less populated areas, has cheaper and flexible deployment \cite{mekki2019comparative}, has an open source gateway, and is robust against multi-path and Doppler effect \cite{vaezi2022cellular}. These facts demonstrated why the LoRa protocol is popular in industry and academia.

Nevertheless, this popular protocol has some \textbf{disadvantages}, such as the lack of QoS and the significant latency, which makes it less attractive for critical applications \cite{mekki2019comparative}. Furthermore, in its v1.0 version, some vulnerabilities can be found during Over The Air Activation (OTAA) \cite{eldefrawy2019formal}. Another aspect is that a LoRa gateway can become a focus for adversaries, as it can be vulnerable to violations \cite{mohamed2022enhancing}.

Additionally, some \textbf{security issues} were listed for LoRaWAN, such as \cite{mohamed2022enhancing} bit flip attack, network flood attack, network traffic analysis, physical attacks, radio frequency (RF) jamming attacks, and self-replay attacks. Other authors \cite{zahariev2022review} complement the list with DoS and DDoS attacks, packet forging, and forged data transmission attacks. As a result, many contributions can be made in the reassessment, adequacy, and correction of vulnerabilities even in this protocol is already in its version 1.1.

\textbf{Narrowband Internet of Things} (\textbf{NB-IoT}) is a protocol proposed by the 3rd Generation Partnership Project (3GPP) in 2015 and standardized in 2016. Moreover, it can use legacy LTE infrastructure as they share a similar architecture or could be deployed as non-cellular \cite{routray2019narrowband}. Additionally, \cite{mekki2019comparative} presents the technology that will serve markets with higher added value and willingness to pay more for low latency and QoS. 

As for \textbf{signal characteristics}, NB-IoT is a narrow band with a band of 200Hz \cite{lalle2019comparative}. Further, devices with this technology can have a battery life of 10 years when sending an average of 200 bytes per day \cite{mekki2019comparative} which is enough for different monitoring applications. Also, it can connect to about 50k end devices per cell, reaching different values for UL (200 kbps) and DL (20 kbps) using SC-FDMA and OFDMA, respectively. As a result, NB-IoT can support a massive number of restricted devices for a long time without much energy consumption.

Furthermore, one of the \textbf{main strengths} of NB-IoT is using the legacy LTE network \cite{rastogi2020narrowband}, which makes it available in several regions, even in rural areas \cite{ugwuanyi2021survey}, as argued by the authors. Likewise, other advantages are deep coverage, low power consumption, low complexity, support of a massive number of connections \cite{rastogi2020narrowband}, and allows a large payload (up to 1600 bytes) \cite{mekki2019comparative}. Also, security is inherited from the LTE network, bringing a strong data transmission security to the system and the end devices. Comparatively, the such protocol has positive results for QoS, latency, and reliability, being the best option in terms of reliability \cite{jouhari2022survey}. Thereupon, as for the cellular network protocols, NB-IoT has the best prospects for future developments and competes directly with LoRaWAN.

Although, this protocol presents some issues to be resolved. Some \textbf{drawbacks} include higher energy consumption by NB-IoT compared to LoRa \cite{thaenkaewevaluating} and Sigfox devices. Also, the base station deployment's high cost and the coverage are shorter than the other two \cite{mekki2019comparative} previews cited. Moreover, for \cite{coman2019security}, this protocol should be used with reservations for critical applications. As a result, NB-IoT is a good option for IoT projects, albeit it is not suitable for every situation.

Regarding Rastogi et al. \cite{rastogi2020narrowband}, some \textbf{open issues} and future evolutions for NB-IoT are time division duplex, mobility management, security, latency, and simulator in NB-IoT. As for \cite{jing2014security}, the transport layer security issues are routing attacks, DoS, and data transit attacks. Lastly, \cite{jha2021layer} presents as threats the replay attack, flooding, black hole attack, and wormhole as vulnerabilities from the network layer. In brief, NB-IoT research and evolution still need to address many vulnerabilities.


Lastly, \textbf{Sigfox} is a protocol designed for low-budget data transmission applications and low-cost, battery-backed devices transmitting data over long distances \cite{anani2019survey}. Further, this protocol was developed in 2010 by the company of the same name \cite{mekki2019comparative} in France. Since its radio chipset costs only two euros and the subscription costs one euro per year, this is a very interesting protocol for developing low-cost projects \cite{lalle2019comparative}. Thus, Sigfox aggregates low power consumption and low-budget hardware offering a low-cost solution for IoT systems.

Due to \textbf{data transmissions}, it happens with a small amount of data, at a low baud rate, and infrequently. Specifically, the data rate is 100 bps and has a limited number of messages transmitted per day (four messages for DL communications and 140 for UL \cite{lalle2019comparative}). Further, there is also a 12 bytes size limitation for the payload of each \cite{mekki2019comparative} transmission, which makes this protocol's suitable for smart metering or similar context . Likewise LoRa, Sigfox operates on the Industrial Scientific and Medical (ISM) band, which has different bands according to the global region (e.g., 433 MHz in Asia, 868 MHz in Europe, and 915 MHz in North America). Besides, It uses an ultra-narrow band (100 Hz) and has a maximum data rate of 100 bps \cite{lalle2019comparative}.

Regarding \textbf{advantages}, this protocol has excellent coverage (>40 km), which means that an entire small city can be served by a single base station \cite{mekki2019comparative}. Also, its low cost is another positive point of this technology, and its radio chipset costs only 2 euro. Since it transmits the same message three times at different frequencies, it is very robust against jamming \cite{lalle2019comparative}. As for security, it implements an end-to-end authentication between the device and the cloud \cite{anani2019survey}. Furthermore, Sigfox provides a platform that supports the collection and storage of packages, providing an API for querying the data. Thereby, this factor is important for companies that want to consume data from IoT sensors but do not have budget to get employees responsible for the network infrastructure.

On the \textbf{downside}, its infrastructure is still available in a few countries \cite{jouhari2022survey}. Thus, even if this protocol is desirable for the application of a project, the lack of Sigfox infrastructure in the country can be an impediment. In addition, this protocol uses its own platform \cite{petrariu2021sigfox}, as described above and it can be a negative factor for companies planning to have greater control over the process. Additionally, it is not suitable for critical applications or applications that need to go beyond measurements, as the number of uploads and its payload is very small (12 bytes) \cite{anani2019survey} not supporting applications that go far beyond latency-tolerant remote monitoring.

Some Sigfox limitations and \textbf{security issues} can be challenging to solutions to adopt this technology. The first is the limitation on the number of messages and the payload size, which makes application layer security solutions suffer restrictions in their implementation. Per second, this protocol may suffer from DoS due to the small size of the Sequence Number (SN) \cite{coman2019security} used to replay attack avoidance. Finally, as it is a bundle solution that incorporates since the radio chip to the cloud platform whereby the system that adopts it become very dependent on the Sigfox company, causing a vendor lock-in problem. In the future, if it is necessary to change the transmission technology, this could impact the entire system having very high costs. Thus, causing an indeterminate dependence on the same technology supplier. Although all three have good performance, the LoRaWAN protocol was selected to be used in the proposal of this work, and due to this, the protocol is presented in more detail.

\subsection{The LoRaWAN}
\label{subsec_lorawan}

As for \textbf{detailing the LoRaWAN protocol}, its \cite{de2021systematic,lin2017using} \textbf{architecture} is composed of End Device, Network Server, Join Server, and Application Server. The End Device is the end device that communicates with the Join Server to authenticate and participate in the network. Once authenticated, it communicates with the Network Server, and the latter forwards the request to the Application Server, which is accessible to the end user. In the case of \cite{jradi2021overview} roaming support, the Network Server can also undergo three specializations: Home Network Server, Serving Network Server, and Forwarding Network Server.

\begin{figure}[h]
    \centering
    \includegraphics[width=0.75\linewidth]{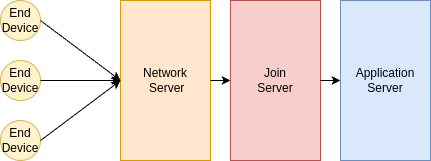}
    \caption{LoRa Archtecture}
    \label{fig:archtecture}
\end{figure}

As for \textbf{deployment}, it can be either public or private \cite{fujdiak2022insights}. In the public deployment, the infrastructure has a large scope and is the responsibility of this or a large operator. In the proven deployment option, the infrastructure usually belongs to a property, company, or university and spans a few kilometers, i.e., a campus.

Devices in LoRa are divided into \textbf{classes A}, \textbf{B}, and \textbf{C} \cite{jouhari2022survey,de2021systematic,ugwuanyi2021survey}. Firstly, \textbf{class A} or All devices have lower energy consumption and have two RX after a TX with a waiting time until the end of the cycle. By the second, the \textbf{class B} or Beacon are the ones that present the behavior of the class A plus with RXs scheduled during the wait, plus a beacon signal. Finally, \textbf{class C}, or Continous, starts with the same behavior as class A but instead of going on hold after two RX, it stays in RX until the end of the cycle. In this way, the first two classes use batteries as an energy source, while the C class needs energy continuously (main powered).

Finally, it is important to discuss LoRaWAN's \textbf{security features} in more detail. Security in LoRaWAN can be splitted into two layers \cite{thaenkaewevaluating}: confidentiality using AES-CMM and integrity through AES-CMAC. Two keys are used during their communication process, the AppKey and the NwkKey root keys that come from the factory on the devices from v1.1 and use AES-128.


Furthermore, after observing the characteristics of LPWAN protocols, it is possible to notice the need for a lightweight security approach in this context. Using mechanisms adopted in networks with greater data transmission capacity is unsuitable in a context with this limitation in the amount of bytes trafficked and with such low transmission speed. However, LPWAN networks usually have low operating costs, low device design costs, and long transmission distances, which make them of paramount importance in the evolution of Smart Metering, specifically in our case. monitoring system (Environmental Sensing). Because of this, the existing security mechanisms should be adapted to the constrains of LPWAN networks.

\section{Discussion}
\label{sec_discussion}

This section will present a general discussion of the results obtained and an analysis of their implications for the final result of the work. Each step that contributed to resolving the general objective of this work is briefly commented on and evaluated here, and some of its main aspects are listed.

As we have seen throughout the work, the main challenges and opportunities in using CD and LPWAN were addressed. Many of the challenges are related to security issues due to the novelty of the related technologies, poor device configuration, and problems in the platforms or networks used, in addition to the already expected problem in the power supply. As opportunities, we have several benefits for producing and conserving natural resources, such as saving water and pesticides and improving the quality of life. Thus, the challenges encountered are characterized more as technological issues that can be easily resolved in the coming years. However, the benefits, in turn, directly impact the quality of life in the long term. As a result, investing in this type of technology for agriculture and forest monitoring, among others, is an investment in the quality of life in the future.

Several important pieces of information were brought up to analyze LPWAN protocols. The oldest technology analyzed is SigFox, which is about 15 years old. All of them use AES-128, but only NB-IoT uses a licensed band. Sigfox has the worst transmission rate but the best coverage distance. They all support many devices ($>$=50k), but NB-IoT supports twice as many as the others. As a result, after analyzing different combined factors, mainly regarding security aspects and openness of the technologies, we chose the LoRaWAN protocol to continue the evaluation.

\begin{table}[ht]
    \centering
    \scalebox{.65}{
    \begin{tabular}{lll}
    \toprule
    \textbf{Protocol} & \textbf{Article} & \textbf{Threats} \\
    \midrule
        \textbf{LoRaWAN} & \cite{mohamed2022enhancing} & \makecell[l]{Bit flip attack,\\ Network flood attack,\\ Network traffic analysis,\\ Physical attacks,\\ Radio frequency (RF) jamming attacks,\\ and Self-replay attacks.} \\
        \cmidrule(l){2-3}
         & \cite{zahariev2022review}& \makecell[l]{DoS and DDoS attacks,\\ Packet forging,\\ and Forged data transmission attacks.}\\
        \midrule
        \textbf{NB-IoT} & \cite{rastogi2020narrowband} & \makecell[l]{Do not support complex security \\algothims\\Eavesdroping\\Malicius nodes attack}\\
        \cmidrule(l){2-3}
         & \cite{jing2014security} & \makecell[l]{Routing attacks,\\ DoS,\\ and Data transit attacks.}\\
        \cmidrule(l){2-3}
         & \cite{jha2021layer} & \makecell[l]{Replay attack,\\ Flooding,\\ Black hole attack,\\ and Wormhole as vulnerabilities from\\ the network layer.} \\
        \midrule
        \textbf{SigFox} & \cite{coman2019security} & \makecell[l]{Limitation on the number of messages\\ and the payload size;\\DoS due to the small size \\of the Sequence Number;\\ Replay attack;\\ Vendor lock-in.} \\        
        \bottomrule
    \end{tabular}
    }
    \caption{Protocol threat list}
    \label{tab:protocol_threat_list}
\end{table}

No less important is the set of threats that are exposed by technology. Some articles present the main threats for each type of data transmission technology, Table \ref{tab:protocol_threat_list}. We can group them into technology limitations, such as the Sigfox limitation on the number of messages, MITM attacks, physical and signal attacks, and network attacks. Although there are vulnerabilities, they can be circumvented and mitigated because they are already well known.

Continuing with the LoRaWAN protocol selected in the previous stage, a slightly more specific view of the protocol was presented, where the architecture of a LoRa network and security aspects already embedded in this technology were discussed. A classification of LoRa devices according to their class was also provided. As a result, together with the previous phase of the analysis, this phase complements the set of information necessary to confirm the LoRaWAN protocol as the most suitable in our perspective for use in sparse remote sensor communication networks or smart agriculture.

\section{Conclusion}
\label{sec_conclusion}

Thus, we conclude the work by evaluating LPWAN protocols for monitoring sparsely distributed sensors. As contributions, a set of challenges and opportunities (Table \ref{tab:challenges_opportunities}) in this area of monitoring using LPWAN and CD were presented. Furthermore, a qualitative analysis of the protocols in several dimensions was presented, Table \ref{tab:lpwan_comparative_qualitative}, and a set of threats that may affect each one, Table \ref{tab:protocol_threat_list}. Finally, a detailed description of the LoRaWAN protocol was provided, and the protocol was selected as the best solution for this scenario.

In future work, other studies should be done on the protocol's security, its network and energy performance, and its integration with other forms of data transmission (e.g., satellite). A more in-depth study in each of these areas would provide a more current and mature view of this technology's future developments, especially for IoT.

\section{Acknowledgments}
The authors sincerely thank the Federal University of Santa Catarina (UFSC). This study was partially funded by the Fundação de Amparo à Pesquisa e Inovação do Estado de Santa Catarina (FAPESC), Edital 20/2024.

\bibliographystyle{cas-model2-names}
\bibliography{cas-refs}

\end{document}